\documentclass[aps,twocolumn,prd,superscriptaddress,nofootinbib]{revtex4}
\usepackage[dvips]{graphicx}
\usepackage{amsmath}
\usepackage{amssymb}
\newcommand{\beq}{\begin{equation}}
\newcommand{\eeq}{\end{equation}}
\newcommand{\bea}{\begin{eqnarray}}
\newcommand{\eea}{\end{eqnarray}}
\newcommand{\ave}[1]{\langle {#1} \rangle}
\newcommand{\eq}[1]{Eq.~(\ref{#1})}
\newcommand{\eqs}[1]{Eqs.~(\ref{#1})}
\begin{document}

\title{A note on color neutrality in NJL-type models}


\author{Michael Buballa}
\email{michael.buballa@physik.tu-darmstadt.de}
\affiliation{Institut f\"ur Kernphysik, 
Technische Universit\"at Darmstadt,
D-64289 Darmstadt, Germany}

\author{Igor A.\ Shovkovy}
\email{shovkovy@th.physik.uni-frankfurt.de}
  \altaffiliation[on leave from ]{%
       Bogolyubov Institute for Theoretical Physics,
       03143, Kiev, Ukraine}
\affiliation{
Frankfurt Institute for Advanced Studies, J.W.\ Goethe-Universit\"at,
D-60438 Frankfurt am Main, Germany}


\date{\today}

\begin{abstract}
By referring to the underlying physics behind the color charge neutrality
condition in quark matter, we discuss how this condition should be
properly imposed in NJL-type models in a phenomenologically meaningful
way. In particular, we show that the standard assumption regarding
the use of two color chemical potentials, chosen in a very special
way, is not justified in general. When used uncritically, such an
approach leads to wrong or unphysical conclusions.
\end{abstract}

\maketitle

It is generally expected that cold dense quark matter is a color 
superconductor.
(For reviews on color superconductivity, see Ref.~\cite{reviews}.)
At asymptotically high densities, when the QCD running coupling 
constant is very small, this can be shown from first principles
using weak-coupling techniques
\cite{weak}. 
At not-so-large densities, however, as present, e.g., in the 
cores of compact stars, the weak-coupling expansion breaks down.
In order to get at least some schematic picture of the phase 
structure in this regime, one therefore has to invoke models. 
In this context, NJL-type models play a prominent role.
(For a recent review, see Ref.~\cite{Buballa:2003qv})
These models are characterized by the fact that the gluon fields
of QCD are replaced by point-like quark-quark or quark-antiquark
interactions. 

In general, the interactions are chosen to be
consistent with the most important symmetries of QCD.
However, due to the lack of gauge fields, the local $SU(3)$-color 
($SU(3)_c$)
symmetry of QCD becomes replaced by a global one in the NJL model.
This simplification has important consequences:

In QCD, the homogeneous color superconducting ground state is automatically 
color neutral. According to Gauss' law, any sample of 
color-charged matter would create a color-electric field, leading to a 
diverging energy density in the infinite volume limit.
Thus, matter is forced to remain color-neutral
by the generation of a color-electrostatic potential,
i.e., a non-zero gluon condensate $\langle A^0_a\rangle\neq 0$
in one or more of the eight color components $a$ \cite{AlRa02}.
(Note, however, that here and in the entire article we assume that the 
matter remains homogeneous, while we cannot exclude the emergence of
locally colored inhomogeneous solutions, like mixed phases \cite{mix}
or LOFF phases \cite{cryst}, as long as they are globally color neutral.)

More recently, the above arguments have been confirmed by calculating
gluon condensates within the framework of QCD at asymptotic
densities \cite{GeRe03,Kry03,DiRi04}.
For instance, for the standard set-up of a two-flavor color superconductor
(see \eq{stans} below), one finds $\langle A^0_8\rangle\neq 0$
and $\langle A^0_a\rangle = 0$ in all other components. 
It is obvious that such a condensate is well defined only after gauge fixing
and breaks the remaining global color symmetry spontaneously.
In particular, this implies that it could be rotated into other color 
directions by global gauge transformations.\footnote{Employing a {\it local} 
gauge transformation, it could even be eliminated completely. However, 
in the corresponding gauge the diquark condensate would pick up a 
space-time dependent color phase, making it rather inconvenient for 
practical purposes.}

In NJL-type models, on the other hand, there are no gauge fields which 
can realize the color neutrality dynamically.
As a consequence, the color-superconducting ground state is not 
automatically color neutral. 
In order to get a physically meaningful result, one therefore has to
impose color neutrality by hand \cite{AlRa02,IiBa01,notneut}. 
To this end, one introduces a set of color chemical potentials
$\{\mu_a\}$ which are chosen in such a way that the corresponding
color charge densities vanish:
\beq
    n_a = \ave{\psi^\dagger T_a \psi} =
    -\frac{\partial\Omega}{\partial\mu_a} = 0~.
\label{neutcond}
\eeq
Here
\beq
    \psi = \left(\begin{array}{c} 
                 \psi_r \\ \psi_g \\ \psi_b \end{array}\right)
\eeq
is a quark field with three color components (``red'', ``green'',
``blue''), $T_a = \frac{\lambda_a}{2}$ denote the generators
of $SU(3)_c$, and $\Omega = -\frac{T}{V} \ln {\cal Z}$ is the grand canonical 
potential of the system per volume.

In the light of the above discussion, the color chemical potentials should
be identified with the corresponding components of the gluon condensate in 
QCD, $\mu_a \equiv g\,\ave{A_a^0}$. 
In fact, from the general path-integral expression of the partition function
one obtains
\beq
    n_a = \frac{1}{2} {\rm Tr} \left[S(x,x)\,\Gamma_a^0\right]~,
\label{tad}
\eeq
where $S$ is the dressed (Euclidean) Nambu-Gorkov quark propagator and
\beq
    \Gamma_a^0 = \frac{\partial S^{-1}}{\partial \mu_a} = 
    \left( \begin{array}{cc} \gamma^0\,T_a & 0 \\
                        0 & -\gamma^0\,T_a^T \end{array} \right)~.
\eeq
is the quark-gluon vertex in Nambu-Gorkov space.
Thus, the density $n_a$ takes the form of a tadpole, corresponding to
a gluon field $A^0_a$ attached to a quark loop.

From this perspective it appears natural to consider eight different
color chemical potentials, $a = 1, \dots, 8$, and to impose \eq{neutcond}
for the corresponding eight derivatives. 
This is at variance with the ``standard definition'' of color neutrality
in NJL-type models as having equal densities of red, green and blue quarks,
\beq
    n_r = n_g = n_b~.
\label{rgb}
\eeq
Here $n_c = \ave{\psi_c^\dagger \psi_c}$. This condition translates into 
requiring \eq{neutcond} only for the diagonal generators,
$a = 3$ and $a = 8$, 
\beq
    n_3 = \frac{1}{2}(n_r-n_g) = 0 \;\; \text{and} \;\;
    n_8 = \frac{1}{2\sqrt{3}}(n_r+n_g-2n_b) = 0~, 
\label{n38}
\eeq
while the non-diagonal generators are ignored.\footnote{Non-diagonal
color densities and the corresponding chemical potentials have been
considered in the early Ginzburg-Landau analysis in Ref.~\cite{IiBa01},
but they got practically ``forgotten'' in most later works.}

However, keeping in mind the underlying physics, it is not always 
a priori clear whether the restriction of \eq{neutcond} to $n_3$ and 
$n_8$ is justified.
In fact, since both, the gluon condensates $\langle A^0_a \rangle$
and the ``tadpoles'' \eq{tad} are subject to gauge transformations,
it is obvious that it cannot be physically correct under all circumstances. 
For instance, one can easily rotate $n_3$ and $n_8$ away by an
appropriate global $SU(3)_c$ transformation. Then,
since such a transformation should leave the physics invariant, if the
original state was not color neutral, the transformed state cannot
be color neutral either, even if both, the new $n_3$ and the new $n_8$, 
vanish. In that case there must be non-vanishing non-diagonal 
density components and their disregard may lead to false conclusions.  
As we will discuss, this is exactly what happened in two recent papers
\cite{He:2005jq,Blaschke:2005km} on the ground state of color neutral 
two-flavor color superconductors (2SC).

To this end, let us consider the specific example of
a two-flavor system in the 2SC phase.
In the standard ansatz for this phase, only the red and green quarks
participate in a diquark condensate, whereas the blue quarks remain
unpaired:
\begin{alignat}{1}
    \Delta_3 \sim& \ave{\psi^T C\gamma_5 \tau_2 T_2 \psi} \neq 0~,
\nonumber\\
    \Delta_2 \sim& \ave{\psi^T C\gamma_5 \tau_2 T_5 \psi} = 0~,
\nonumber\\
    \Delta_1 \sim& \ave{\psi^T C\gamma_5 \tau_2 T_7 \psi} = 0~.
\label{stans}
\end{alignat}
Here $\tau_2$ is a Pauli matrix acting is flavor space, while the 
$T_a$ are the generators of $SU(3)_c$, as before. 
If we use a common chemical potential $\mu$ for all colors, 
the resulting ground state is not color neutral, but the densities of
the quarks participating in the condensate, i.e., the red and green 
ones, are larger than the density of the blue quarks:
\beq
    n_r = n_g > n_b \quad \Rightarrow \quad n_8 > 0~.
\label{n8g0}
\eeq
It turns out that this is the only non-vanishing color density 
for this state \cite{DiRi04} (see also below). 

In this situation, the standard procedure is to introduce a 
color chemical potential $\mu_8$ which is chosen in such a way that $n_8$
vanishes.
As a result, the free energy gets somewhat larger, i.e., the 
color neutral ground state is formally less favored than
the colored one at $\mu_8=0$. 
Of course, this should be considered as an artifact of the 
NJL model: In QCD, the presence of long-range color forces would lead
to a diverging energy density and therefore prevent the emergence
of homogeneous colored solutions.  
Instead, as pointed out above, a static gluon field arises which
exactly neutralizes the color charge. 
Therefore, in the NJL model, the color-neutral NJL-model solution 
which is obtained after introducing a non-vanishing $\mu_8$ should be
considered as the physical one. 

However, it was recently claimed in Ref.~\cite{He:2005jq} that 
this solution is unstable with respect to the emergence of non-zero
condensates $\Delta_1$ and $\Delta_2$. This was seemingly confirmed
by Ref.~\cite{Blaschke:2005km} who found 
a symmetric state with $\Delta_1 = \Delta_2 = \Delta_3$
to be the most favored ``color neutral'' solution.
In both references, color neutrality was defined as in \eq{n38}, 
i.e., restricted to the diagonal color charges. 

This result is easy to understand.
Since the origin of the non-vanishing $n_8$ in
\eq{n8g0} is the fact that the blue quarks remain unpaired in the
standard ansatz \eq{stans}, it is plausible that a symmetric ansatz,
where all colors are paired in a ``democratic'' way, should lead to 
equal densities of red, green and blue quarks without introducing a 
non-zero $\mu_8$.  
On the other hand, the symmetric ansatz can be obtained from \eq{stans} 
by a rotation in color space.
Thus, since the NJL model Lagrangian is invariant under global $SU(3)$
transformations, the rotated ground state is degenerate with
the colored ground state of the standard ansatz \eq{stans} at $\mu_8 = 0$.
Recalling that the latter is lower in free energy than the color neutral 
solution with non-zero $\mu_8$, we could naively conclude that we have
found a new color neutral solution that is more favored than the standard
one.

Of course, this cannot be true: The rotated
ground state is physically equivalent to the unrotated one and
therefore it cannot be color neutral. This means, if $n_3 = n_8 = 0$ 
for the symmetric ansatz, its color content must be hidden in the 
non-diagonal components.    

In order to analyze this in more detail, we consider a continuous 
$SU(3)_c$ transformation that rotates the conventional 
diquark condensate $(0,0,\Delta_3)$ into $(\Delta_1',\Delta_2',\Delta_3')$
with $\Delta_1'=\Delta_2'$.
This is given by
\beq
    \psi \;\rightarrow\; \psi' = U\,\psi~,
\label{psirot}
\eeq
with 
\begin{alignat}{1}
    U &= \exp\left[i\varphi(\lambda_5-\lambda_7)\right]
\nonumber\\[1mm]
      &=
\left( \begin{array}{ccc}
               \phantom{-}\cos^2\left(\frac{\varphi}{\sqrt{2}}\right) &
               \phantom{-}\sin^2\left(\frac{\varphi}{\sqrt{2}} \right)& 
               \phantom{-}\frac{\sin(\sqrt{2}\varphi)}{\sqrt{2}}\\[2mm]
               \phantom{-}\sin^2\left(\frac{\varphi}{\sqrt{2}}\right) & 
               \phantom{-}\cos^2\left(\frac{\varphi}{\sqrt{2}}\right) &
                        -\frac{\sin(\sqrt{2}\varphi)}{\sqrt{2}} \\[2mm]
                        -\frac{\sin(\sqrt{2}\varphi)}{\sqrt{2}} &
               \phantom{-}\frac{\sin(\sqrt{2}\varphi)}{\sqrt{2}} &
               \phantom{-}\cos(\sqrt{2}\varphi)
               \end{array} \right).
\label{rotmat}
\end{alignat}
The diquark condensates are then transformed as
\begin{alignat}{1}
    &\ave{\psi^T C\gamma_5 \tau_2 T_a \psi} 
\nonumber\\
     &\quad \rightarrow\;
     \ave{{\psi'}^T C\gamma_5 \tau_2 T_a \psi'} \;=\;   
     \ave{\psi^T C\gamma_5\,\tau_2\,U^T T_a U \psi}~.
\end{alignat}
Indeed,
\begin{alignat}{1}
&U^{T}T_2 U  = \cos(\sqrt{2}\varphi) \,T_2 \,-\,
             \frac{1}{\sqrt{2}}\sin(\sqrt{2}\varphi)\,\left(T_5 + T_7 \right)~,
\nonumber\\
&U^{T}T_5 U  = 
\nonumber\\ 
&\quad       \frac{1}{\sqrt{2}} \sin(\sqrt{2}\varphi) \,T_2 
             \,+\,\cos^2(\frac{\varphi}{\sqrt{2}})\,T_5 
             \,-\,\sin^2(\frac{\varphi}{\sqrt{2}})\,T_7~,
\nonumber\\
&U^{T}T_7 U  = 
\nonumber\\ 
&\quad       \frac{1}{\sqrt{2}} \sin(\sqrt{2}\varphi) \,T_2 
             \,-\,\sin^2(\frac{\varphi}{\sqrt{2}})\,T_5 
             \,+\,\cos^2(\frac{\varphi}{\sqrt{2}})\,T_7~,
\end{alignat}
which yields for $\Delta_1=\Delta_2=0$:
\beq
    \Delta_1'=\Delta_2'= \frac{1}{\sqrt{2}}\sin(\sqrt{2}\varphi)\,\Delta_3~,
    \qquad
    \Delta_3'= \cos(\sqrt{2}\varphi)\,\Delta_3~.
\label{Dprime}
\eeq
In particular, this implies the symmetric case 
\beq
    \Delta_1' = \Delta_2' = \Delta_3' = \frac{\Delta_3}{\sqrt{3}}\quad
    \text{for} \quad \cos(\sqrt{2}\varphi) = \frac{1}{\sqrt{3}}~.
\label{Dsym}
\eeq

By using the same color transformation, one could introduce the following 
set of two generators of the Cartan algebra:
\bea
T_3^\prime & \equiv & U^{\dagger} T_3 U \nonumber \\
&=& \cos(\sqrt{2}\varphi)\, T_3 
+\frac{1}{\sqrt{2}}\sin(\sqrt{2}\varphi)\,\left(T_4+T_6\right) 
\nonumber\\
T_8^\prime & \equiv & U^{\dagger}\, T_8 U \nonumber\\
&=& \frac{1}{4}\left(1+3\cos(2\sqrt{2}\varphi)\right)\, T_8
+\frac{\sqrt{3}}{2} \sin^2(\sqrt{2}\varphi)\, T_1
\nonumber\\
&&+\frac{1}{2}\sqrt{\frac{3}{2}}\sin(2\sqrt{2}\varphi)\,\left(T_4-T_6\right)~.
\label{Tprime}
\eea
The other generators $T_a$ can be transformed in the analogous way. 
From these relations one can immediately read off the transformed 
color densities
$n_a = \ave{{\psi'}^\dagger T_a \psi'}$.
For the interesting case that $n_3$ and $n_8$ are the only non-vanishing
$n_a$ in the unrotated state one finds:
\begin{alignat}{1}
  n_1' &= \frac{\sqrt{3}}{2} \sin^2(\sqrt{2}\varphi)\, n_8~, 
\nonumber \\
  n_3' &= \cos(\sqrt{2}\varphi)\, n_3~, 
\nonumber \\
  n_4' &= -\frac{1}{\sqrt{2}}\sin(\sqrt{2}\varphi)\,n_3
          -\frac{1}{2}\sqrt{\frac{3}{2}}\sin(2\sqrt{2}\varphi)\,n_8~,
\nonumber \\
  n_6' &= -\frac{1}{\sqrt{2}}\sin(\sqrt{2}\varphi)\,n_3~
          +\frac{1}{2}\sqrt{\frac{3}{2}}\sin(2\sqrt{2}\varphi)\,n_8~,
\nonumber \\
  n_8' &= \frac{1}{4}\left(1+3\cos(2\sqrt{2}\varphi)\right)\,n_8~,
\label{nprime}
\end{alignat}
and $n_2' = n_5' = n_7' = 0$.
Note that this implies that the length $|\vec n| = \sqrt{\sum n_a^2}$
of the vector $\vec n = (n_1, \dots, n_8)$ remains invariant under 
color rotations. 

For the symmetric case, \eq{Dsym}, one finds
\begin{alignat}{1}
  n_1' &= \frac{1}{\sqrt{3}}\, n_8~,
\nonumber \\
  n_3' &= \frac{1}{\sqrt{3}}\, n_3~,
\nonumber \\
  n_4' &= -\frac{1}{\sqrt{3}}\,(n_3+n_8)~, 
\nonumber \\
  n_6' &= -\frac{1}{\sqrt{3}}\,(n_3-n_8)~,
\end{alignat}
and $n_2' = n_5' = n_7' = n_8' = 0$.
Hence, for the 2SC phase with vanishing $\mu_a$, in which
$n_3 = 0$ and only $n_8 \neq 0$ in 
the unrotated state, \eq{stans}, we indeed find that $n_3'$ and
$n_8'$ vanish in the symmetric state (as well as $n_2'$, $n_5'$, and
$n_7'$), but
\beq
    n_1' = -n_4' = n_6' = \frac{n_8}{\sqrt{3}} \neq 0~.
\eeq
This means that the symmetric ansatz, \eq{Dsym}, without color chemical
potentials does {\it not} lead to a color neutral solution,
in agreement with our general arguments.

Note that the ground state of Ref.~\cite{He:2005jq} can also be obtained
from the conventional configuration in \eq{stans} by a color rotation
as defined in \eqs{psirot} and (\ref{rotmat}) with an infinitesimally
small angle $\varphi$. 
If done properly, one then has to rotate the chemical potentials as well.
Since the $\mu_a$ transform exactly as the $n_a$, it follows from \eq{nprime}
that infinitesimal small chemical potentials $\mu_a$, $a=1,4,6$, are needed
(in addition to $\mu_8$) in order to keep the rotated state color neutral. 
In turn, the neglect of these additional chemical potentials induces 
color charges which have been overlooked in Ref.~\cite{He:2005jq}.
Of course, when correctly neutralized, the rotated state is exactly
degenerate with the unrotated one.

In conclusion, we have pointed out the importance to ensure 
\eq{neutcond} for {\it all} color components $a = 1, \dots, 8$ 
in order to obtain a physically meaningful 
description of color-superconducting phases in NJL-type models.
We have demonstrated that, when used uncritically, the standard approach 
of restricting \eq{neutcond} to $a = 3$ and $8$ could lead to wrong
results. 
This is in line with the underlying physical picture, where the 
various color chemical potentials $\mu_a$ correspond to the time
components of a static gluon field, which transform as members of
an octet under $SU(3)_c$.  

The following remark is in order here. Out of the eight generators,
$T_1, \dots, T_8$, there are only two which are mutually commuting,
namely $T_3$ and $T_8$. This means that there are only two color charges
which can be measured simultaneously. It is obviously convenient to 
choose $n_3$ and $n_8$, but this is not mandatory. It is easy to show
that the rotated operators, $T_3'$ and $T_8'$ (see \eq{Tprime}),
also commute with each other, $[T_3^\prime , T_8^\prime]=0$.
This means that the corresponding color charges, which are linear
combinations of the various components of the $n_a$ in the original basis,
are as ``physical'' as $n_3$ and $n_8$. 
This is similar to the more familiar case of angular momentum: While it
is customary to measure $J_3$, i.e., to choose the $z$-axis as quantization 
axis, one could equally well quantize the system along any other axis.

Also note that for any combination of $n_a$,
one can always find a color rotation with 
\beq
    (n_1, \dots , n_8) \rightarrow 
    (0,0,n_3',0,0,0,0,n_8')~,
\eeq
i.e., it is always possible to work in a basis where $n_3$ and $n_8$
are the only non-vanishing components (at most). 
In a {\it given} basis, however, the restriction to $n_3$ and
$n_8$ without detailed analysis is in general not justified.
   
Finally, it is natural to ask whether the restriction of \eq{neutcond} 
to $a = 3$ and $8$ has already led to wrong conclusions in other cases
in the literature.
In particular, one should ask whether
starting from \eq{stans} and tuning the value of $\mu_8$ is sufficient
to achieve color neutrality in the 2SC phase. 
To answer this question, we can directly employ \eq{tad}.
For the (gapless) 2SC phase, the explicit form of the propagator $S$
has been determined, see, e.g.,  Refs.~\cite{weak,Huang:2004am}.
Its color structure contains only two types of terms, namely
$diag_c(1,1,0)$ and $diag_c(0,0,1)$. Thus, only their traces with
$T_8$ can be nontrivial, i.e., there can be no non-zero $n_a$, 
except of $n_8$. 
It can be shown in a similar way that only $T_3$ and $T_8$
tadpoles appear in the (gapless) CFL and uSC phases \cite{Stefan}.
Moreover, it has been shown in Ref.~\cite{Kry03} that only $A_3^0$
and $A_8^0$ condense in the CFL$+K^0$ phase.
Hence, at least for the most common phases, the standard procedure
turns out to be sufficient. It is important, however, to verify this
statement case by case.

\begin{acknowledgments}
The authors would like to thank V. Miransky, K. Rajagopal, D.~Rischke, 
M. Ruggieri, and A.~Schmitt for interesting discussions and useful comments.
This work was supported in part by the Virtual Institute of the
Helmholtz Association under grant No. VH-VI-041 and by Gesellschaft
f\"{u}r Schwerionenforschung (GSI) and by Bundesministerium f\"{u}r
Bildung und Forschung (BMBF).
\end{acknowledgments}



\end{document}